\documentclass{article}
\usepackage{spconf,amsmath,graphicx}
\usepackage{booktabs}
\usepackage{multirow}
\usepackage{tikz}
\usepackage{pgfplots}
\usepackage{hyperref}
\usepackage{cite}
\usepackage{caption}
\pgfplotsset{width=6cm, compat=1.4}

\title{ZERO RESOURCE CODE-SWITCHED SPEECH BENCHMARK USING SPEECH UTTERANCE PAIRS FOR MULTIPLE SPOKEN LANGUAGES}
\name{Kuan-Po Huang$^{1*}$, Chih-Kai Yang$^{2*}$, \thanks{*Equal contribution.} Yu-Kuan Fu$^3$, Ewan Dunbar$^4$, Hung-yi Lee$^5$}
\address{
$^{1235}$National Taiwan University \quad $^{1}$ASUS Intelligent Cloud Services \quad $^{4}$University of Toronto
}
\begin{document}
\ninept
\maketitle
\setlength{\abovedisplayskip}{3pt}
\setlength{\belowdisplayskip}{3pt}
\begin{abstract}
We introduce a new zero resource code-switched speech benchmark designed to assess the code-switching capabilities of self-supervised speech encoders directly. We showcase a baseline system of language modeling on discrete units to demonstrate how the code-switching abilities of speech encoders can be assessed in a zero-resource manner. Our experiments encompass a variety of well-known speech encoders, including Wav2vec 2.0, HuBERT, XLSR, etc., on three tracks of different code-switched language pairs: Spanish-English, French-English, and Chinese-English. We examine the impact of pre-training languages and model size on benchmark performance. Notably, though our results demonstrate that speech encoders with multilingual pre-training, exemplified by XLSR, outperform monolingual variants (Wav2vec 2.0, HuBERT) in code-switching scenarios, there is still substantial room for improvement in their code-switching linguistic abilities.

\end{abstract}
\vspace{-10pt}
\section{Introduction}
\label{sec:intro}
Code-switching is a common phenomenon in our daily lives, especially in conversations between people from different regions or countries with multiple official languages. In speech processing, there are also various kinds of tasks where code-switching might be involved, for example, speech recognition \cite{liu2023reducing, hamed2023benchmarking}, speech translation \cite{weller-etal-2022-end, huber2022codeswitching}, text-to-speech synthesis \cite{zhao2020natural}, etc. With the huge advantage of using heavily parameterized self-supervised speech encoders such as Wav2vec 2.0 \cite{baevski2020wav2vec}, HuBERT \cite{hsu2021hubert} and its variants \cite{lee2022textless, huang22b_interspeech, huang2023_ekd, huang2023_impgen}, and XLSR \cite{conneau2020unsupervised, babu2021xlsr}, many of the speech processing tasks are performed on the representations extracted by these speech encoders. Thus, code-switching abilities become essential for their applicability to tasks involving code-switching. However, to our best knowledge, no existing benchmark or corpus allows the speech community to directly evaluate the inherent code-switching abilities of these commonly used speech encoders. Hence, we propose a zero resource code-switched speech benchmark to address this issue.

To achieve direct assessment, we designed a task requiring syntactic and semantic abilities of code-switching and adopted spoken language modeling systems as our baseline systems.
The advantages of directly assessing the code-switching ability of speech encoders in a zero-shot manner are twofold: one is that additional parameters of downstream models are not needed, and the other one is that paired downstream training data and labels are not required. This not only relieves the burden of training multiple downstream models when there are many downstream tasks but also allows us to utilize unlabeled speech data to serve as the training data during the assessment process instead of having to collect paired training data, which is extremely difficult in the code-switching scenario.

Our newly proposed task requires the model to distinguish correct speech utterances from wrong ones. More specifically, given a pair of code-switched utterances, with one being normal and the other making no sense, meaningless, or grammatically unacceptable, the model has to prefer the normal one over the erroneous one by assigning a higher score to it. A speech encoder would have to attain both semantic and syntactic linguistic abilities in a code-switching scenario to obtain good results on this newly proposed metric. Based on this task, we constructed a benchmark with testing tracks of code-switched spoken language pairs, including Spanish-English, French-English, and Chinese-English.

Speaking of code-switching, knowing that code-switching involves more than one language in a sentence, there has been a debate on whether multilingual text-based LLMs have code-switching abilities \cite{yong2023prompting, zhang2023multilingual, winata2021multilingual}. Relatively, we also looked into the code-switching ability of multilingual self-supervised speech encoders. Unfortunately, our results indicated that the evaluated speech models still have a long way to go regarding code-switching ability.

Overall, the contributions of our zero resource code-switched speech benchmark are: (1) Proposing a new zero resource code-switched speech task for assessing syntactic and semantic linguistic abilities of self-supervised speech models in code-switching scenarios, (2) Highlighting that there is significant room for improvement for several existing multilingual speech models in such a task. 

Data samples and code of our baseline systems are available at \url{https://github.com/nobel861017/cs_zs_baseline}.

\vspace{-10pt}
\section{Related work}
\label{sec:related}
\textbf{BLiMP and sBLIMP.} The Benchmark of Linguistic Minimal Pairs (BLiMP) \cite{warstadt-etal-2020-blimp-benchmark} in the Natural Language Processing field is a task with pairs of sentences, where each pair consists of one grammatically correct sentence while the other one is ungrammatical. This task aims to evaluate the linguistic ability of text-based language models by investigating whether the models can assign a higher probability to the grammatically correct sentence. Later on, \cite{nguyen2020zero} proposed a zero resource speech benchmark, including a speech version of BLiMP, namely, sBLIMP. Similar to BLiMP, this task also contains pairs of sentences but in the form of speech. The key difference between the baseline systems of sBLIMP and BLiMP is that the former takes discrete units quantized from speech representations as input while the latter takes text as input. The goal of sBLIMP is to evaluate the syntactic ability of speech encoders, while BLiMP is to evaluate the syntactic ability of text-based language models. 

\textbf{Spoken language modeling for linguistic assessment.} The Zero Resource Speech Challenge 2021 \cite{nguyen2020zero} established a baseline system demonstrating how speech encoders could be evaluated through spoken language modeling directly based on speech without the need for textual information. They made use of the intrinsic capabilities of each component in the system to assess the linguistic abilities of the speech encoders in several different aspects. Especially for probability-based metrics like sBLIMP, they adopted the spoken language model to assign probability scores to the utterances in each pair and interpreted the resulting sentence acceptability accuracy as how well the syntactic understanding of the speech encoder is. In this way, the assessment could be completed without textual resources and labeled speech data, and thus, direct assessment could be achieved. In our work, we adopted a similar methodology and extended the logic to a novel code-switching task.

\vspace{-5pt}

\section{Zero resource code-switched speech task}
\label{sec:zero-cs}

Our proposed zero resource code-switched speech task is similar to sBLIMP. Each data pair consists of two spoken utterances, a correct one and a wrong one. The goal is to assign a higher score to the correct utterance. Slightly different from the previous works, the term ``correct" in this scenario means that the content of an utterance makes sense and is meaningful and grammatically acceptable. 

Take the correct and wrong sentence in the lower block of Fig. \ref{fig:cs_prompt} for example. To understand the correct sentence, the system should have multilingual knowledge. Specifically, it needs to have English ability to understand what ``water" is and Chinese ability to know the Chinese part of the sentence means ``This does not dissolve in something." Furthermore, cross-lingual understanding is necessary for it to incorporate its semantic understanding in the two languages to know that the sentence means ``This does not dissolve in water.". Similarly, the system should use multilingual and cross-lingual capabilities to understand the other sentence as ``This does not dissolve in fire". Finally, as the first one is more meaningful, the assigned score should be higher than that assigned to the other one.

This example shows that to achieve good performance on the proposed task, the model needs multilingual and cross-lingual syntactic and semantic understanding in a code-switching scenario. Thus, we expect this task to provide a way to assess the code-switching linguistic abilities of the self-supervised speech models.

We note that there are many linguistic theories of code-switching that attempt to explain, among other things, why some grammatical positions are impossible for code-switching \cite{sankoff1981formal,myers2017code}. While some of the incorrect sentences are indeed grammatically inappropriate (as confirmed by our human evaluations in Section \ref{data-gen}), our benchmark does require us to have an answer to when code-switching is grammatically allowed. In many cases, the incorrect sentence simply generates semantic incoherence. Nevertheless, the benchmark measures a model's ability to do language modeling in the presence of code-switching.

\vspace{-10pt}

\subsection{Data generation and validation}
\label{data-gen}
\begin{figure}[t]
    \centering
    \includegraphics[width=8.5cm]{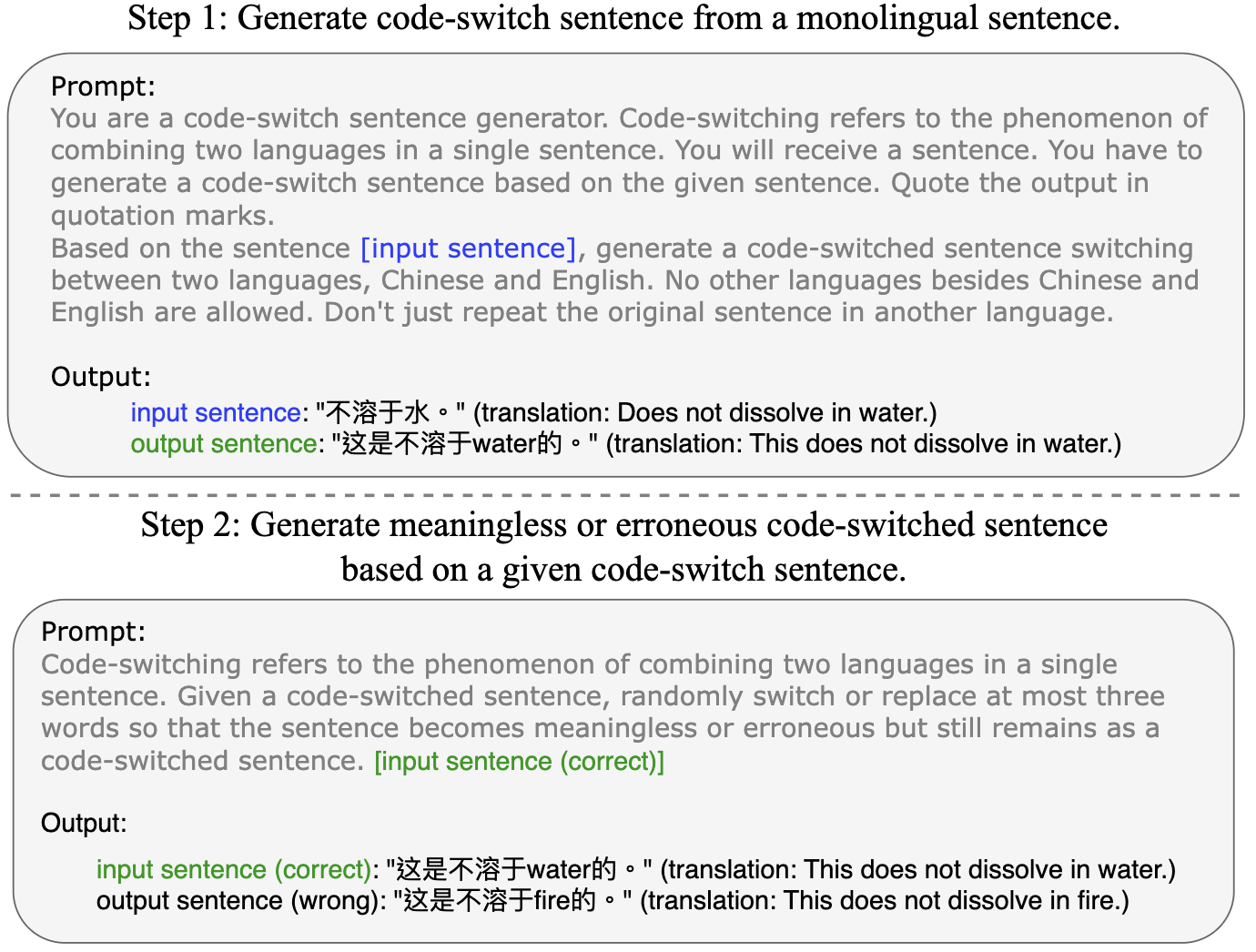}
    \vspace{0.5em}
    \setlength{\belowcaptionskip}{-15pt}
    \caption{Code-switched text data generation by prompting ChatGPT.}
    \label{fig:cs_prompt}
\end{figure}

To generate pairs of correct and wrong utterances, we first utilized the well-known LLM released by OpenAI, ChatGPT, to generate code-switched sentences in which English (en) is mixed with either Spanish (es), French (fr), or Chinese (zh). As shown in Fig. \ref{fig:cs_prompt}, we prompted ChatGPT by first defining code-switching as suggested in \cite{yong2023prompting} and asking it to generate a code-switched sentence based on a given monolingual sentence in language $X$ from Common Voice \cite{ardila2019common}, where $X \in \{\text{es, fr, zh}\}$, to restrict the content of the resulting sentence to some extent (Step 1 in Fig. \ref{fig:cs_prompt}). The generated sentence with English mixed with language $X$ would be used as the presumed correct sentence, and the corresponding wrong sentence was generated by requiring ChatGPT to replace or switch at most three words in the presumed correct sentence so that the resulting sentence could be more meaningless or erroneous than the original one (Step 2 in Fig. \ref{fig:cs_prompt}) while preserving the overall similarity between the two sentences. We discovered that the wrong sentences generated in this way actually tended to make no sense, be meaningless, or get grammatically unacceptable. Finally, to synthesize the code-switched speech pairs, we adopted the Amazon Polly system \cite{amazonpolly} to synthesize bilingual speech utterances.
As suggested in \cite{yong2023prompting}, we conducted human validation by multiple bilingual speakers. Each human annotator was required to label whether the paired sentences were valid, meaning that the presumed correct sentence in each pair should (1) actually make sense and be meaningful and grammatically acceptable, (2) be indeed better than the presumed wrong one in the aspects above. Pairs failing to meet the above two requirements would be labeled invalid. To ensure the annotation quality and consistency, the hired annotators were required to complete an annotation trial on some sampled paired sentence data with pre-defined ground truths. Human annotators were required to get at least $95\%$ accuracy before proceeding to the data annotating process. A pair of correct and wrong sentences was included in the task if most of the annotators labeled it valid.

\vspace{-7pt}
\subsection{Code-switched data statistics}
This zero resource code-switched speech benchmark has three tracks based on three code-switched language pairs, including Spanish-English (es-en), French-English (fr-en), and Chinese-English (zh-en), with 7263, 4020, and 3176 human-validated data samples, respectively. For each language pair, all available bilingual speaker configurations were adopted from the Amazon Polly text-to-speech system to synthesize utterances, and the configurations of the two utterances in each pair were the same. All the synthesized utterances had a sample rate of 22.5kHz originally and were later resampled to 16kHz to match the configurations of the speech encoders.

\vspace{-10pt}

\subsection{Baseline systems}
\begin{figure*}[ht]
    \centering
    \includegraphics[width=16.0cm,height=3.2cm]{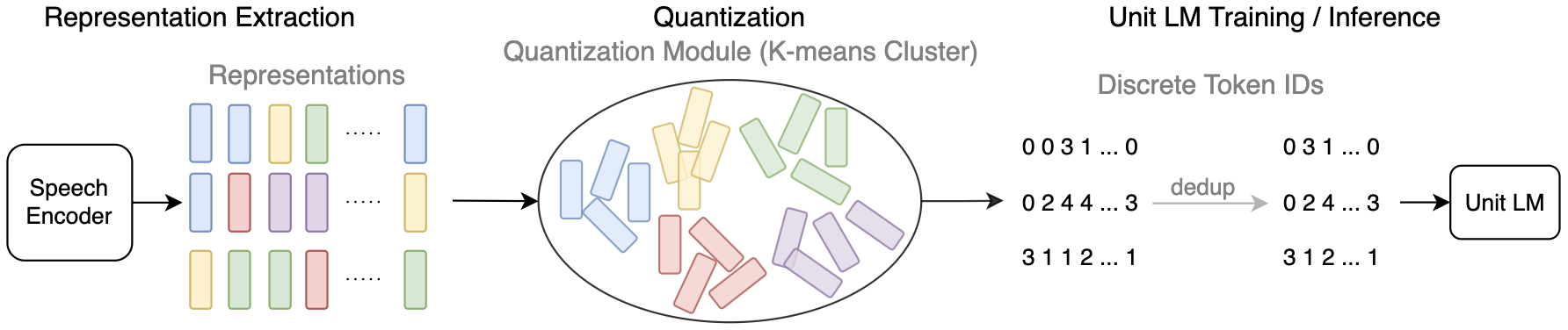}
    \vspace{0.5em}
    \setlength{\belowcaptionskip}{-13pt}
    \caption{Illustration of our speech-based baseline systems with discrete unit language modeling.}
    \label{fig:zr_speech_pipe}
\end{figure*}

Our speech-based baseline systems are depicted in Fig. \ref{fig:zr_speech_pipe}, which follow the design of a typical spoken language modeling system and consist of three main modules: the speech encoder, the quantization module, and the unit language model (Unit LM). Given a speech dataset, representations of part of the dataset are first extracted by the speech encoder and are then formed into $k$ clusters via the k-means algorithm. The resulting k-means clusters will further serve as the quantization module for the whole training split of the speech dataset. For each representation, the quantization is done by assigning the ID of the cluster the representation vector belongs to, and thus, the originally continuous waveforms become sequences of discrete units. Following previous works using speech units \cite{lee2022direct, lee2022textless}, after the quantization of the whole training split, a deduplication operation is performed to ensure that there are no successive identical units in the unit sequences. Note that this operation is not performed in the original spoken language modeling system in \cite{nguyen2020zero}. Finally, the collected unit sequences are used as the training data to train the Unit LM. After the training, the testing set is discretized by the quantization module, and the Unit LM is used to compute the probabilities (span-PP score mentioned in Section \ref{score}) of the correct and wrong utterances of the pairs for evaluation.

For reference, we provide some direct-inference results of pre-trained text-based language models from fairseq \cite{ott-etal-2019-fairseq}, including XLM-R \textsc{Base} \cite{conneau2020XLMR} and XGLM 1.7B \cite{lin2021few}. We also include a random baseline derived by utilizing random-assigned units and a random-weighted Unit LM to simulate the performance of the baseline systems in a totally random situation.

\vspace{-9pt}

\subsection{Evaluation metric}
\label{score}
The performance of this code-switched speech task is measured in accuracy, where a hit occurs when the Unit LM assigns a higher span-masked pseudo-probability (span-PP) score \cite{nguyen2020zero} to the correct utterance. Given a discrete unit sequence of a quantized speech utterance $\mathbf{u} = u_1, u_2, \cdots, u_T$, the span-PP score is defined as follows:
\begin{equation}
\begin{aligned}
    & \text{span-PP}_{w, s} (\mathbf{u})\\ =& \prod_{i = 1 + j\cdot s} P(u_i \cdots u_{i+w}|u_1\cdots u_{i-1}u_{i+w+1}\cdots u_T)
\end{aligned}
\end{equation}
where $w$ is the decoding span size, $s$ is the stride, and $0 \leq j \leq\ \lfloor (T-1)/s \rfloor$. In our work, $w$ and $s$ are set to be 15 and 5, respectively.



\vspace{-5pt}
\section{Experimental setup}
\label{sec:exp}
\subsection{Training set}
\label{subsec:train_set}
The training sets in our experiments were sampled from the following speech corpora: LibriSpeech \cite{panayotov2015librispeech} for English (en), Multilingual LibriSpeech \cite{pratap2020mls} for Spanish (es) and French (fr), and MAGICDATA Mandarin Chinese Read Speech Corpus \cite{magicdata} for Chinese (zh). Note that as our experiments aimed to assess the inherent code-switching ability of the pre-trained multilingual and monolingual speech encoders and served as the baselines of the benchmark, we didn't use any code-switched data for training to prevent potential learning of code-switching abilities from those data and the possible bias in the resulting performance. 
\vspace{-6pt}
\subsection{Speech encoders, Quantization modules, and Unit LMs}
\textbf{Speech encoders.} In our baselines, we picked several widely-used pre-trained speech models publicly available at fairseq \cite{ott-etal-2019-fairseq} and S3PRL \cite{s3prl}, including XLS-R 1B, XLS-R 0.3B \cite{babu2021xlsr}, XLSR-53 \cite{conneau2020unsupervised}, Wav2vec 2.0 \textsc{Large} \cite{baevski2020wav2vec}, HuBERT \textsc{X-Large}, HuBERT \textsc{Base} \cite{hsu2021hubert}, and mHuBERT \cite{lee2022textless} as the speech encoders to investigate if they can solve a code-switching task even though code-switched data were absent during pre-training. 

As the generalizability to the code-switching task of these models and the underlying relationship between such abilities and the layers of the models remain unexplored, in our baselines, only the hidden representations of the last layer of the encoders were extracted for the training of the quantization module and the discretization of the dataset. We leave the layer-wise analysis of these models' performance on the proposed task as future work.

\textbf{Quantization modules.} For the quantization modules required in our baseline systems, we sampled monolingual data from the speech corpora mentioned in Section \ref{subsec:train_set}, forming different sets of training data for each speech encoder. Each set resulted in 100 hours of monolingual speech in total and consisted of the languages the corresponding speech encoder had seen during its pre-training phase. For each speech encoder, a k-means model with $k=100$ was trained with its corresponding set of monolingual speech and served as the quantization module by assigning the ID numbers of the closest cluster centers to the vectors at each time step. 

\textbf{Unit LMs.} Similar to the training data of the quantization modules, we sampled monolingual data of the languages involved in the pre-training of the speech encoders and formed a training set containing 400 hours in total. The training set was further discretized with the quantization modules to obtain the training set for the Unit LMs. We then trained BERT \textsc{Base} models on the discretized training set to serve as the Unit LM, with the masked token prediction as the training objective. Following \cite{baevski2020vqwav2vec} and \cite{nguyen2020zero}, spans of $M$ consecutive tokens were masked for the model to predict, where $M \sim \mathcal{N}(10, 100)$. The training was done with a total batch size of 2.6M tokens, and the learning rate was warmed up to the peak value of $10^{-4}$ and polynomially decayed afterward. The implementation was based on fairseq \cite{ott-etal-2019-fairseq}.
\vspace{-13pt}
\section{Results}
\label{sec:results}
\vspace{-5pt}
\begin{table*}[ht]
\setlength\tabcolsep{5.0 pt}
\renewcommand{\arraystretch}{0.4}
\caption{Performance of the speech encoders, text-based models, and the random baseline in accuracy (\%) on es-en, fr-en, and zh-en tracks. 
}
\centering
\begin{tabular}{lcccccccc}
\toprule
           & \# param.   & km: 100 cluster        & Unit LM (RoBERTa)       &  & es-en  & fr-en  & zh-en  & avg \\
Speech encoder         & (B) & mono speech (hr)           & mono speech (hr)            & dedup        & Acc $\uparrow$   & Acc $\uparrow$   & Acc $\uparrow$ & Acc $\uparrow$  \\
\midrule
\midrule
\multicolumn{9}{c}{Multilingual Speech Encoders}\\
\midrule
XLSR-53 (53 lang)               & 0.3       & es, fr, zh, en 25 each & es, fr, zh, en 100 each & V            & 33.74 & 45.25 & 47.20 & 42.06 \\
XLS-R 0.3B  (128 lang)           & 0.3       & es, fr, zh, en 25 each & es, fr, zh, en 100 each & V            & 75.16 & 59.30 & 43.18 & 59.21\\
XLS-R 1B     (128 lang)          & 1         & es, fr, zh, en 25 each & es, fr, zh, en 100 each & V            & 33.30 & 38.66  & 39.22 & 37.06\\
mHuBERT (es, fr, en)               & 0.09      & es, fr, en 33 each     & es, fr, en 133 each     & V            & 29.55 & 30.42 & 40.33 & 33.43 \\
\midrule
\midrule
\multicolumn{9}{c}{Monolingual Speech Encoders}\\
\midrule
Wav2vec 2.0 LARGE (ll60k) & 0.3       & en 100                 & en 400                  & V            & 13.11 & 25.35 & 42.41 & 26.96\\
HuBERT X-LARGE (ll60k)        & 1         & en 100                 & en 400                  & V            & 24.54 & 25.60 & 38.60 & 29.58\\
HuBERT Base (LS960)           & 0.09      & en 100                 & en 400                  & V            & 22.26 & 25.30 & 40.24 & 29.27 \\
\midrule
\midrule
random                 & -         & random                 & random                  & -            & 23.63 & 32.11 & 37.47 & 31.07\\
\midrule
\midrule
XLM-RoBERTa Base (text-base)&  0.125 & - & - & - & 54.62 & 55.12	& 55.16 & 54.97\\
XGLM 1.7B (text-base) &  1.7 & - & - & - & 90.91 &	88.38 & 92.03 & 90.44\\
\bottomrule
\end{tabular}
\label{tab:dedup}
\vspace{-12pt}
\end{table*}

The overall results are listed in Table. \ref{tab:dedup}, with the number of pre-training languages of multilingual speech encoders, the corpora the monolingual speech encoders were pre-trained on, and the number of parameters of these encoders included for reference.

Please note that although we tried our best to restrict the length difference of the sentences to balance the length of the utterances between the correct and wrong versions, it's still impossible to make the lengths exactly matched for every pair. Thus, there may be a bias in favor of the shorter utterance, which was generally the wrong one. While most of the speech-based baselines were not significantly influenced, perhaps because of their informative units, and could apparently distinguish the two utterances as the span-PP scores of the two utterances differed a lot, the random baseline was heavily misled since its units were randomly assigned, making its performance fall below 50\%. Nevertheless, this only shifts the reference level for comparison, and the interpretation of the results remains the same.
\vspace{-10pt}
\subsection{Multilingual pre-training}
\vspace{-5pt}
Comparing the results of the baseline systems with multilingual speech encoders (the uppermost block in Table \ref{tab:dedup}) and those with monolingual ones (the middle block in Table \ref{tab:dedup}), it is obvious that the systems with multilingual speech encoders substantially outperform those with their monolingual counterparts in es-en and fr-en tracks. 

As for the zh-en track, except for XLS-R 1B, all the models that included Chinese during pre-training slightly outperform their monolingual counterpart (Wav2Vec2.0 \textsc{Large}), though the differences are insignificant. This may be due to relatively inadequate pre-training data in Chinese compared with those in Spanish and French. For mHuBERT, the performance on the zh-en track is close to that of HuBERT \textsc{Base} since Chinese was absent during pre-training.

Overall, the results show that multilingual pre-training does help in the proposed task and serves as evidence that our benchmark can effectively distinguish the models' multilingual abilities. 

\vspace{-11pt}

\subsection{Model size and pre-training languages}
\vspace{-5pt}
Comparing the performance of baseline systems with XLSR-53, XLS-R 0.3B, and XLS-R 1B in Table \ref{tab:dedup}, we first observe that systems with XLSR-53 and XLS-R 0.3B as speech encoders consistently outperform that with XLS-R 1B in all the tracks, even though these two models have much fewer parameters than XLS-R 1B has. However, we do not observe a similar trend in the comparison between systems trained with HuBERT \textsc{Base} and with HuBERT \textsc{X-Large}. This suggests that the model with a smaller size may extract representations that have a stronger capability of generalizing to a task requiring out-of-domain code-switching knowledge, but such an advantage will conditionally appear if the model meets the minimal requirements of the abilities needed to solve the task (multilingual ability, in this case).

Next, we find that the system with XLS-R 0.3B significantly outperforms that with XLSR-53, which may imply that the multilingual pre-training with broader coverage of languages provides better generalizability for code-switching and thus induces better performance on our benchmark. 

Note that these two observations are similar to those discovered in \cite{shi23g_interspeech}. As XLS-R 0.3B benefits from both model size and the broad coverage of pre-training languages, the baseline system based on it achieves the best performance among all the speech-based baselines.

\vspace{-10pt}

\subsection{Deduplication}
\vspace{-5pt}
Comparing the performance of each XLSR model without unit deduplication to their corresponding counterparts with unit deduplication in Table \ref{tab:xlsr_w_wo_dedup}, we find that deduplication always benefits performance on the es-en and fr-en track. In contrast, performance degradation is observed in the zh-en set.
The reason for this degradation requires further investigation in the future.
However, by considering the average performance of the three testing sets, the deduplication operation is still helpful in improving the overall performance of this task.

\vspace{-8pt}
\begin{table}[ht]
\setlength\tabcolsep{4.5pt}
\renewcommand{\arraystretch}{0.4}
\caption{Ablations studies of deduplication for XLSR models. 
}

\centering
\begin{tabular}{lccccc}
\toprule
Speech encoder        & dedup & es-en  & fr-en  & zh-en & avg   \\
\midrule
XLSR-53 (53 lang)     & X        & 32.27 & 42.19 & \textbf{49.65} & 41.37 \\
XLS-R 0.3B (128 lang) & X        & 68.87 & 50.87 & \textbf{44.21} & 54.65 \\
XLS-R 1B (128 lang)   & X        & 29.57 & 35.30  & \textbf{41.91} & 35.59 \\
\midrule
XLSR-53 (53 lang)     & V        & \textbf{33.74} & \textbf{45.25} & 47.20 & \textbf{42.06} \\
XLS-R 0.3B (128 lang) & V        & \textbf{75.16} & \textbf{59.30}  & 43.18 & \textbf{59.21} \\
XLS-R 1B (128 lang)   & V        & \textbf{33.30} & \textbf{38.66} & 39.22 & \textbf{37.06}\\
\bottomrule
\end{tabular}
\label{tab:xlsr_w_wo_dedup}
\vspace{-20pt}
\end{table}

\subsection{Gap between speech-based and text-based systems}
\vspace{-5pt}
The lowermost block of Table \ref{tab:dedup} shows the performance of evaluating text-based language models on the transcripts of the testing set. We find that the pre-trained XLMR \textsc{Base}, which has the same architecture as all the Unit LMs of the speech-based baselines and has been pre-trained on a large amount of multilingual data, can not obtain satisfactory performance, indicating that this task is not easy for a multilingual text-based model with moderate size. The task is difficult because it requires faithful encoding of not only the phonetics but also the semantic and grammatical properties of words in two different languages. 
However, even this unsatisfactory performance outperforms most of our speech-based baselines built on commonly used speech encoders that have been reported to be powerful in several downstream tasks. This implies that this task is even harder for existing speech encoders. We also notice that there is a tremendous gap between the best performance of speech-based baselines and the text-based models. This is likely due to the overall limitations of unit quality in current systems, which also affect the performance of monolingual language modeling on previous monolingual syntactic (sBLIMP) and semantic evaluations in the Zero Resource Speech Challenge \cite{dunbar2021zero}. In sum, this suggests that there is still room for these speech models to improve on this code-switching task and hence, on the code-switching syntactic and semantic abilities.

\vspace{-10pt}
\section{Conclusion}
\label{sec:conclusion}
\vspace{-7pt}
This paper introduces a novel benchmark to assess the code-switching capability of self-supervised speech models in a zero-shot manner. Our results show that the size of speech models and the coverage of pre-training languages considerably influence the models' generalization ability for this out-of-domain code-switching task. In addition, the results unveil that most of the evaluated speech models do not exhibit strong code-switching ability compared to the text-based language models and still have a long way to go. We invite the speech community to participate in this benchmark and encourage further research on broadening the speech processing technology for code-switching. 

\section{Acknowledgement}

We thank the National Center for High-performance Computing (NCHC) of the National Applied Research Laboratories (NARLabs) in Taiwan for providing computational and storage resources.
\bibliographystyle{IEEEbib}
\bibliography{strings,refs}

\begin{thebibliography}{10}

\bibitem{liu2023reducing}
Hexin~Liu et~al.,
\newblock ``Reducing language confusion for code-switching speech recognition with token-level language diarization,''
\newblock in {\em ICASSP 2023}.

\bibitem{hamed2023benchmarking}
Injy Hamed and Amir~Hussein et~al.,
\newblock ``Benchmarking evaluation metrics for code-switching automatic speech recognition,''
\newblock in {\em SLT}. IEEE, 2022, pp. 999--1005.

\bibitem{weller-etal-2022-end}
Orion Weller and Matthias~Sperber et~al.,
\newblock ``End-to-end speech translation for code switched speech,''
\newblock in {\em Findings of ACL}, Dublin, Ireland, May 2022, pp. 1435--1448, Association for Computational Linguistics.

\bibitem{huber2022codeswitching}
Christian Huber, Enes~Yavuz Ugan, and Alexander Waibel,
\newblock ``Code-switching without switching: Language agnostic end-to-end speech translation,''
\newblock {\em arXiv:2210.01512}, 2022.

\bibitem{zhao2020natural}
Shengkui Zhao, Trung~Hieu Nguyen, Hao Wang, and Bin Ma,
\newblock ``{Towards Natural Bilingual and Code-Switched Speech Synthesis Based on Mix of Monolingual Recordings and Cross-Lingual Voice Conversion},''
\newblock in {\em INTERSPEECH 2020}.

\bibitem{baevski2020wav2vec}
Alexei Baevski, Henry Zhou, Abdelrahman Mohamed, and Michael Auli,
\newblock ``wav2vec 2.0: a framework for self-supervised learning of speech representations,''
\newblock in {\em Proceedings of the 34th International Conference on Neural Information Processing Systems}, 2020, pp. 12449--12460.

\bibitem{hsu2021hubert}
Wei-Ning~Hsu et~al.,
\newblock ``Hubert: Self-supervised speech representation learning by masked prediction of hidden units,''
\newblock {\em TASLP}, vol. 29, pp. 3451--3460, 2021.

\bibitem{lee2022textless}
Ann Lee, Hongyu Gong, Paul-Ambroise Duquenne, Holger Schwenk, Peng-Jen Chen, Changhan Wang, Sravya Popuri, Yossi Adi, Juan Pino, Jiatao Gu, et~al.,
\newblock ``Textless speech-to-speech translation on real data,''
\newblock in {\em NAACL}, 2022, pp. 860--872.

\bibitem{huang22b_interspeech}
Kuan~Po Huang, Yu-Kuan Fu, Yu~Zhang, and Hung yi~Lee,
\newblock ``{Improving Distortion Robustness of Self-supervised Speech Processing Tasks with Domain Adaptation},''
\newblock in {\em INTERSPEECH 2022}.

\bibitem{huang2023_ekd}
Kuan~Po Huang, Tzu-Hsun Feng, Yu-Kuan Fu, Tsu-Yuan Hsu, Po-Chieh Yen, Wei-Cheng Tseng, Kai-Wei Chang, and Hung-Yi Lee,
\newblock ``Ensemble knowledge distillation of self-supervised speech models,''
\newblock in {\em ICASSP 2023}.

\bibitem{huang2023_impgen}
Kuan-Po Huang, Yu-Kuan Fu, Tsu-Yuan Hsu, Fabian~Ritter Gutierrez, Fan-Lin Wang, Liang-Hsuan Tseng, Yu~Zhang, and Hung-yi Lee,
\newblock ``Improving generalizability of distilled self-supervised speech processing models under distorted settings,''
\newblock in {\em SLT 2022}.

\bibitem{conneau2020unsupervised}
Alexis~Conneau et~al.,
\newblock ``{Unsupervised Cross-Lingual Representation Learning for Speech Recognition},''
\newblock in {\em INTERSPEECH 2021}, 2021.

\bibitem{babu2021xlsr}
Arun~Babu et~al.,
\newblock ``{XLS-R: Self-supervised Cross-lingual Speech Representation Learning at Scale},''
\newblock in {\em Proc. Interspeech 2022}, 2022, pp. 2278--2282.

\bibitem{yong2023prompting}
Zheng-Xin~Yong et~al.,
\newblock ``Prompting large language models to generate code-mixed texts: The case of south east asian languages,''
\newblock {\em arXiv:2303.13592}, 2023.

\bibitem{zhang2023multilingual}
Ruochen~Zhang et~al.,
\newblock ``Multilingual large language models are not (yet) code-switchers,''
\newblock {\em arXiv:2305.14235}, 2023.

\bibitem{winata2021multilingual}
Genta Indra~Winata et~al.,
\newblock ``Are multilingual models effective in code-switching?,''
\newblock in {\em Proceedings of the Fifth Workshop on Computational Approaches to Linguistic Code-Switching}, 2021, pp. 142--153.

\bibitem{warstadt-etal-2020-blimp-benchmark}
Alex Warstadt, Alicia Parrish, Haokun Liu, Anhad Mohananey, Wei Peng, Sheng-Fu Wang, and Samuel~R. Bowman,
\newblock ``{BL}i{MP}: The benchmark of linguistic minimal pairs for {E}nglish,''
\newblock {\em TACL}, vol. 8, pp. 377--392, 2020.

\bibitem{nguyen2020zero}
Tu~Anh~Nguyen et~al.,
\newblock ``The zero resource speech benchmark 2021: Metrics and baselines for unsupervised spoken language modeling,''
\newblock in {\em NeuRIPS Workshop on Self-Supervised Learning for Speech and Audio Processing}, 2020.

\bibitem{sankoff1981formal}
David Sankoff and Shana Poplack,
\newblock ``A formal grammar for code-switching,''
\newblock {\em Research on Language \& Social Interaction}, vol. 14, no. 1, pp. 3--45, 1981.

\bibitem{myers2017code}
Carol Myers-Scotton,
\newblock ``Code-switching,''
\newblock {\em The handbook of sociolinguistics}, pp. 217--237, 2017.

\bibitem{ardila2019common}
Rosana~Ardila et~al.,
\newblock ``Common voice: A massively-multilingual speech corpus,''
\newblock in {\em Proceedings of the Twelfth Language Resources and Evaluation Conference}, Marseille, France, May 2020, pp. 4218--4222, European Language Resources Association.

\bibitem{amazonpolly}
``{Amazon Polly},'' \url{https://aws.amazon.com/polly/}.

\bibitem{lee2022direct}
Ann Lee, Peng-Jen Chen, Changhan Wang, Jiatao Gu, Sravya Popuri, Xutai Ma, Adam Polyak, Yossi Adi, Qing He, Yun Tang, Juan Pino, and Wei-Ning Hsu,
\newblock ``Direct speech-to-speech translation with discrete units,''
\newblock in {\em ACL}, Dublin, Ireland, May 2022, pp. 3327--3339, ACL.

\bibitem{ott-etal-2019-fairseq}
Myle Ott, Sergey Edunov, Alexei Baevski, Angela Fan, Sam Gross, Nathan Ng, David Grangier, and Michael Auli,
\newblock ``fairseq: A fast, extensible toolkit for sequence modeling,''
\newblock in {\em NAACL (Demonstrations)}, Minneapolis, Minnesota, June 2019, pp. 48--53, Association for Computational Linguistics.

\bibitem{conneau2020XLMR}
Alexis~Conneau et~al.,
\newblock ``Unsupervised cross-lingual representation learning at scale,''
\newblock in {\em ACL 2020}. ACL.

\bibitem{lin2021few}
Xi~Victoria~Lin et~al.,
\newblock ``Few-shot learning with multilingual generative language models,''
\newblock in {\em EMNLP}, Abu Dhabi, United Arab Emirates, Dec. 2022, pp. 9019--9052, Association for Computational Linguistics.

\bibitem{panayotov2015librispeech}
Vassil Panayotov, Guoguo Chen, Daniel Povey, and Sanjeev Khudanpur,
\newblock ``Librispeech: an asr corpus based on public domain audio books,''
\newblock in {\em ICASSP}. IEEE, 2015, pp. 5206--5210.

\bibitem{pratap2020mls}
Vineel~Pratap et~al.,
\newblock ``{MLS: A Large-Scale Multilingual Dataset for Speech Research},''
\newblock in {\em INTERSPEECH 2020}.

\bibitem{magicdata}
``{MAGICDATA},'' \url{https://www.openslr.org/68}.

\bibitem{s3prl}
``{S3PRL},'' \url{https://github.com/s3prl/s3prl}.

\bibitem{baevski2020vqwav2vec}
Alexei Baevski, Steffen Schneider, and Michael Auli,
\newblock ``vq-wav2vec: Self-supervised learning of discrete speech representations,''
\newblock in {\em ICLR}. 2020, OpenReview.net.

\bibitem{shi23g_interspeech}
Jiatong~Shi et~al.,
\newblock ``{ML-SUPERB: Multilingual Speech Universal PERformance Benchmark},''
\newblock in {\em INTERSPEECH 2023}.

\bibitem{dunbar2021zero}
Ewan Dunbar, Mathieu Bernard, Nicolas Hamilakis, Tu~Anh Nguyen, Maureen de~Seyssel, Patricia Rozé, Morgane Rivière, Eugene Kharitonov, and Emmanuel Dupoux,
\newblock ``{The Zero Resource Speech Challenge 2021: Spoken Language Modelling},''
\newblock in {\em INTERSPEECH 2021}.

\end{thebibliography}

\end{document}